\documentclass[11pt]{article}
\usepackage[a4paper,margin=1in,top=25mm]{geometry}
\usepackage{amsmath,amssymb,amsthm}
\usepackage{graphicx}
\usepackage{hyperref}
\usepackage{authblk}
\usepackage{dcolumn}
\usepackage{bm}

\title{GaAs/AlAs Acoustic Nanocavities for Coherent GHz–THz Phonon Engineering}
\author[1]{S. Sandeep}
\author[1]{E. R. Cardozo de Oliveira}
\author[1]{E. Mehdi}
\author[1]{N. D. Lanzillotti-Kimura\thanks{Corresponding author: daniel.kimura@cnrs.fr}}

\affil[1]{Université Paris-Saclay, C.N.R.S., Centre de Nanosciences et de Nanotechnologies (C2N),  10 Boulevard Thomas Gobert, 91120 Palaiseau, France}

\begin{document}

\maketitle

\begin{abstract}
The controlled confinement of high-frequency acoustic phonons in semiconductor nanostructures has emerged as a key ingredient for functional nanophononic and hybrid quantum technologies. In this Review, we summarize recent advances that have established GaAs/AlAs acoustic nanocavities as a versatile and scalable platform for GHz--THz phonon engineering. Compared with alternative nanophononic platforms, GaAs/AlAs offers a particularly favorable combination of mature epitaxial growth, strong photoelastic coupling, and simultaneous optical–acoustic mode colocalization across the GHz–THz regime. We focus on distributed Bragg reflector (DBR)-based architectures, with particular emphasis on micropillar resonators enabling three-dimensional phonon confinement and strong colocalization of acoustic and optical fields. Recent developments in ultrafast optical techniques, including picosecond ultrasonics and Brillouin scattering, have provided unprecedented access to phonon dynamics, coherence, and dissipation at the nanoscale. These advances, combined with strong optophononic coupling, have enabled efficient coherent generation, detection, and manipulation of confined acoustic modes. We discuss key performance metrics, integration strategies, and remaining challenges, notably in acousto-optic transduction efficiency and scalable electrical control. Finally, we outline near-term perspectives for nonlinear phononics, hybrid quantum systems, and integrated phononic circuits, positioning GaAs/AlAs heterostructures as a robust and scalable platform for next-generation nanophononic functionalities.
\end{abstract}

\section{Introduction}
The controlled confinement of high-frequency acoustic phonons in semiconductor nanostructures has recently crossed a threshold from proof-of-concept demonstrations toward functional nanophononic devices. This transition is driven by growing demands in ultrafast data processing \cite{stillerCoherentlyRefreshingHypersonic2020, merklein_chip_2020,Kristensen_Optomechanical_Memory_2024}, sensing \cite{Pan_PhononLaser_2024,Chang_DetectingNanoparticles_2023}, quantum technologies \cite{anderson_two-color_2018, Zhu_optoacoustic_entanglement_2024}, and nanoscale thermal management \cite{Florez_hypersonic_phonon_transport_2022, Braun_thermal_transport2022}. In this context, GaAs/AlAs heterostructures have emerged as a benchmark material platform, combining large acoustic impedance contrast, mature epitaxial growth, and strong optoacoustic coupling to enable distributed Bragg reflector (DBR)-based phononic devices with precise control over acoustic confinement and efficient photon–phonon interactions \cite{fainsteinStandingOpticalPhonons2001,trigoPlanarCavityPrl2002,anguianoMicropillarResonatorsOptomechanics2017}. Together, these advances mark a qualitative transition in nanophononics, where confined high-frequency acoustic phonons evolve from experimental probes into functional degrees of freedom that can be engineered, integrated, and actively controlled within solid-state platforms.

Nanophononics—the study and manipulation of phonons at the nanoscale—shares deep conceptual and architectural analogies with photonics \cite{fainsteinSimultaneousConfinementPRL2013}. In photonics, DBR-based microcavities have enabled transformative advances in nonlinear optics, polariton physics \cite{sunBoseEinsteinCondensationLongLifetime2017,DengCondensationofSemiconductor2002,tawaraCavityPolaritonsInGaN2004,st-jeanLasingTopologicalEdge2017}, high-resolution spectroscopy, and quantum light sources \cite{Somaschi2016,paulSinglephotonEmission1552017,drawerMonolayerBasedSinglePhotonSource2023}. These developments provide a powerful conceptual framework for nanophononics, where analogous cavity architectures are now being exploited to confine, manipulate, and coherently control acoustic phonons at gigahertz and terahertz frequencies.

The foundations of this field were established by two key technological advances. First, progress in epitaxial growth and nanofabrication—initially driven by nanoelectronics and optoelectronics—enabled the realization of semiconductor superlattices with nanometer-scale precision, leading to early demonstrations of selective phonon transmission \cite{narayanamurtiSelectiveTransmissionHighFrequency1979}. Second, the advent of ultrafast optical techniques, including pump--probe spectroscopy and Brillouin scattering, provided direct access to high-frequency acoustic phonons in both the time and frequency domains \cite{thomsenCoherentPhononGeneration1984,kargarAdvancesBrillouinMandelstam2021}. Together, these tools established the experimental basis for nanoscale phonon engineering.

In recent years, renewed interest in nanophononics has been catalyzed by advances in integrated optomechanics, quantum information science, and nanoscale thermal control. These applications require coherent phonon control at sub-micron length scales, low phonon occupation numbers at cryogenic temperatures, and compatibility with scalable device architectures. As a result, earlier concepts have been revisited and extended toward higher frequencies, stronger confinement, and increased functional integration.

Despite substantial progress, key challenges remain, notably the realization of active and tunable nanophononic devices, efficient electrical transduction of high-frequency phonons, and robust operation in the quantum regime. In this Review, we summarize recent advances that have enabled GaAs/AlAs planar and micropillar nanocavities to evolve into a functional nanophononic platform. We focus on design principles, experimental techniques, and device architectures that define the current state of the art, and outline near-term directions toward integrated, scalable, and hybrid nanophononic systems.

\section{Heterostructures for phonon confinement}
\label{nanostructures}

Phononic crystals, the acoustic analogs of photonic crystals, provide a powerful framework for controlling the propagation of elastic waves in solids. Their potential spans fundamental wave physics, non-destructive characterization, thermal transport, and emerging quantum technologies. While nanophononics has become a highly active field in recent years, early pioneering work already demonstrated the selective transmission of high-frequency acoustic phonons through semiconductor superlattices \cite{narayanamurtiSelectiveTransmissionHighFrequency1979}. Since then, a broad range of nanophononic devices inspired by photonic concepts—including mirrors, cavities, and filters—have been realized using phononic Bragg reflectors \cite{lanzillotti-kimuraNanowaveDevicesTerahertz2006,lanzillotti-kimuraAcousticPhononNanowave2007,trigoPlanarCavityPrl2002,huynhSubterahertzPhononDynamics2006,maznevLifetimeSubTHzCoherent2013a}.

\begin{figure*}
    \includegraphics[width=1\textwidth]{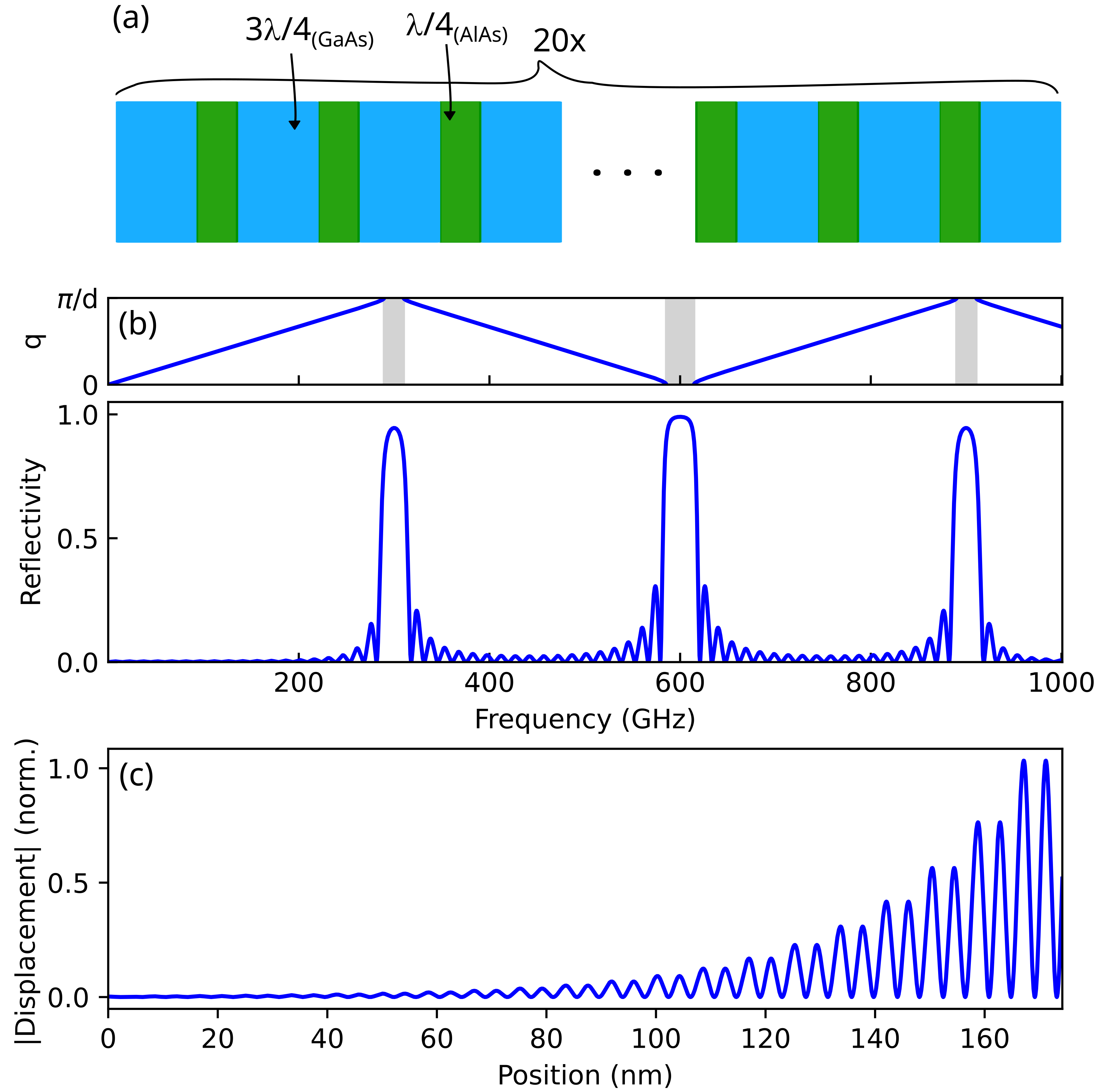}
    \caption{Design principles of DBR-based acoustic phonon confinement in GaAs/AlAs heterostructures. (a) Schematic representation of a GaAs/AlAs superlattice forming a one-dimensional phononic Bragg reflector. (b) Corresponding folded acoustic dispersion relation and reflectivity spectrum, highlighting the formation of stopbands that enable strong phonon reflection. (c) Spatial profile of the acoustic displacement field for a representative confined mode at 600 GHz, illustrating exponential attenuation inside the mirror. This figure summarizes the basic physical principles underlying phonon confinement in DBR-based nanophononic devices.}
    \label{superlattice}
\end{figure*}

In bulk semiconductors, the dispersion relation of longitudinal acoustic phonons is approximately linear over a wide frequency range. Periodic multilayer heterostructures composed of materials with contrasting acoustic impedances induce a folding of the acoustic dispersion relation, giving rise to phononic bandgaps at the Brillouin zone center and edge (Figure~\ref{superlattice}). Finite superlattices thus act as acoustic distributed Bragg reflectors (DBRs), exhibiting high-reflectivity stopbands whose spectral position and width are governed by the acoustic impedance contrast, layer thicknesses, and number of periods. These well-established design principles underpin the operation of all DBR-based nanophononic architectures discussed in the following sections.

GaAs/AlAs superlattices constitute a particularly versatile material system for high-frequency phonon confinement, owing to their large acoustic impedance contrast and compatibility with high-quality epitaxial growth. Optimized DBR designs enable strong reflection for longitudinal acoustic phonons propagating along the growth direction, resulting in exponential attenuation of the displacement field within the mirror for frequencies lying inside the stopbands. Importantly, the quasi-linear phonon dispersion in this material system allows simultaneous operation at multiple stopband orders without significant variation of the elastic properties, facilitating flexible device design across a broad frequency range.

Beyond their role as passive reflectors, GaAs/AlAs superlattices are efficient platforms for the generation and detection of coherent acoustic phonons. Strong photoelastic coupling associated with polariton resonances enables highly sensitive detection schemes in ultrafast pump--probe experiments \cite{kobeckiGiantPhotoelasticityPolaritons2022,karzelPolaritonProbingAttometre2025}. When the layer thickness is reduced to the nanometer scale, forming multiple quantum well structures, additional mechanisms such as piezoelectric coupling to confined carriers further enhance coherent phonon generation \cite{sunCoherentAcousticPhonon2000,wenEfficientGenerationCoherent2007}. These geometries also allow control over the spectral content of the generated phonon wavepackets, including the relative intensities of fundamental and harmonic components \cite{chernSpectralAnalysisHighharmonic2003}. Superlattice-based structures have furthermore been shown to support guided acoustic modes at gigahertz frequencies \cite{yaremkevichProtectedLongDistanceGuiding2021} and to enhance lateral acoustic confinement when integrated with nanowires \cite{manteGigahertzCoherentGuided2013}.

Semiconductor heterostructures can also be actively driven using high-amplitude acoustic waves, enabling dynamic control over electronic and optical properties. Ultrafast strain pulses have been shown to induce electrical currents at frequencies exceeding that of the driving strain waveform, offering routes toward tunable millimeter- and submillimeter-wave sources \cite{wangUltrafastStrainInducedCharge2020}. Acoustic excitation can further modulate excitonic and optical responses, including transient suppression and recovery of excitonic resonances and enhancement of laser emission \cite{akimovPicosecondAcousticsSemiconductor2015}. In addition, strain pulses provide unique probes of complex internal dynamics, such as polariton collapse and strain-induced transparency phenomena \cite{karzelPolaritonInducedTransparencyMultiple2024}.

\begin{figure*}
    \includegraphics[width=1\textwidth]{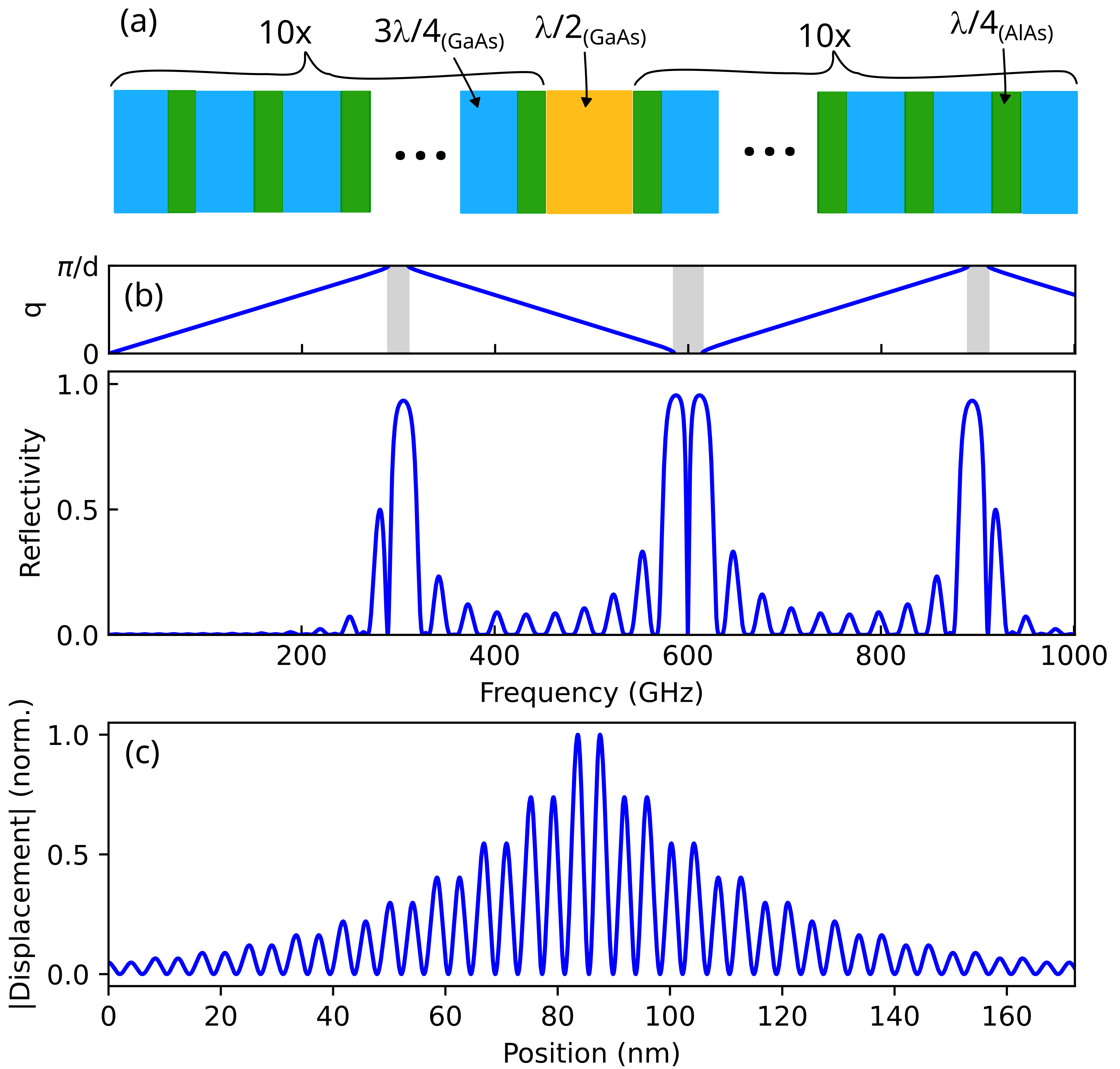}
    \caption{Planar GaAs/AlAs acoustic nanocavity based on distributed Bragg reflectors.
(a) Schematic of a Fabry–Pérot–type phononic cavity formed by two DBRs enclosing a GaAs spacer layer. (b) Acoustic dispersion and reflectivity spectrum, showing a narrow resonance within the stopband associated with a confined cavity mode. (c) Corresponding displacement field profile, with maximum phonon amplitude localized in the spacer. Planar cavities constitute a fundamental building block for high-frequency phonon confinement and optophononic integration.}
    \label{cavity}
\end{figure*}

The simplest confined phononic device derived from these concepts is the acoustic cavity, which represents the nanoacoustic analog of a Fabry--Pérot resonator (Figure~\ref{cavity}). In this geometry, two acoustic DBRs define a spacer layer in which discrete phonon modes are formed when the spacer thickness corresponds to an integer multiple of half acoustic wavelengths \cite{trigoPlanarCavityPrl2002}. The spectral position of the cavity mode is primarily set by the spacer thickness, while the cavity quality factor is determined by the reflectivity of the surrounding DBRs. For typical GaAs/AlAs cavities operating around 600~GHz, total thicknesses on the order of 200~nm are sufficient to achieve quality factors in the range of $10^{3}$--$10^{4}$, providing compact and efficient building blocks for high-frequency nanophononic devices.  While alternative material platforms such as silicon-based heterostructures, diamond, or two-dimensional materials are actively explored for phononic applications, GaAs/AlAs heterostructures currently remain a benchmark system for high-frequency nanophononics. Their unique combination of large acoustic impedance contrast, mature epitaxial growth, and intrinsic optoacoustic coupling enables a level of spectral control, confinement, and hybrid integration that has yet to be matched across the gigahertz–terahertz range.

GaAs/AlAs should not be viewed as a universally superior platform for all nanophononic applications, but rather as one of the most mature and versatile systems for devices requiring simultaneous strong acoustic confinement, optical co-localization, and hybrid quantum functionality. Compared with silicon-based platforms, GaAs/AlAs offers a particularly favorable balance of acoustic impedance contrast and optophononic coupling, enabling highly efficient DBR-based confinement and enhanced optophononic interactions. In contrast to strongly piezoelectric platforms such as LiNbO$_3$ or AlN/GaN, where electrical transduction is often more straightforward, GaAs/AlAs provides superior compatibility with high-finesse optical microcavities, exciton-polariton physics, and embedded quantum emitters such as quantum dots. Diamond and two-dimensional materials offer additional advantages in thermal transport, spin coherence, or extreme mechanical properties, but currently lack the same degree of mature epitaxial control and scalable semiconductor microcavity integration across the GHz–THz regime. In this context, the strength of GaAs/AlAs lies not in maximizing a single figure of merit, but in enabling the simultaneous engineering of optical, acoustic, and electronic functionalities within the same heterostructure platform.

\section{Fabrication: from planar structures to micropillar resonators and waveguides}
\label{fabrication}

\begin{figure*}
    \includegraphics[width=1\textwidth]{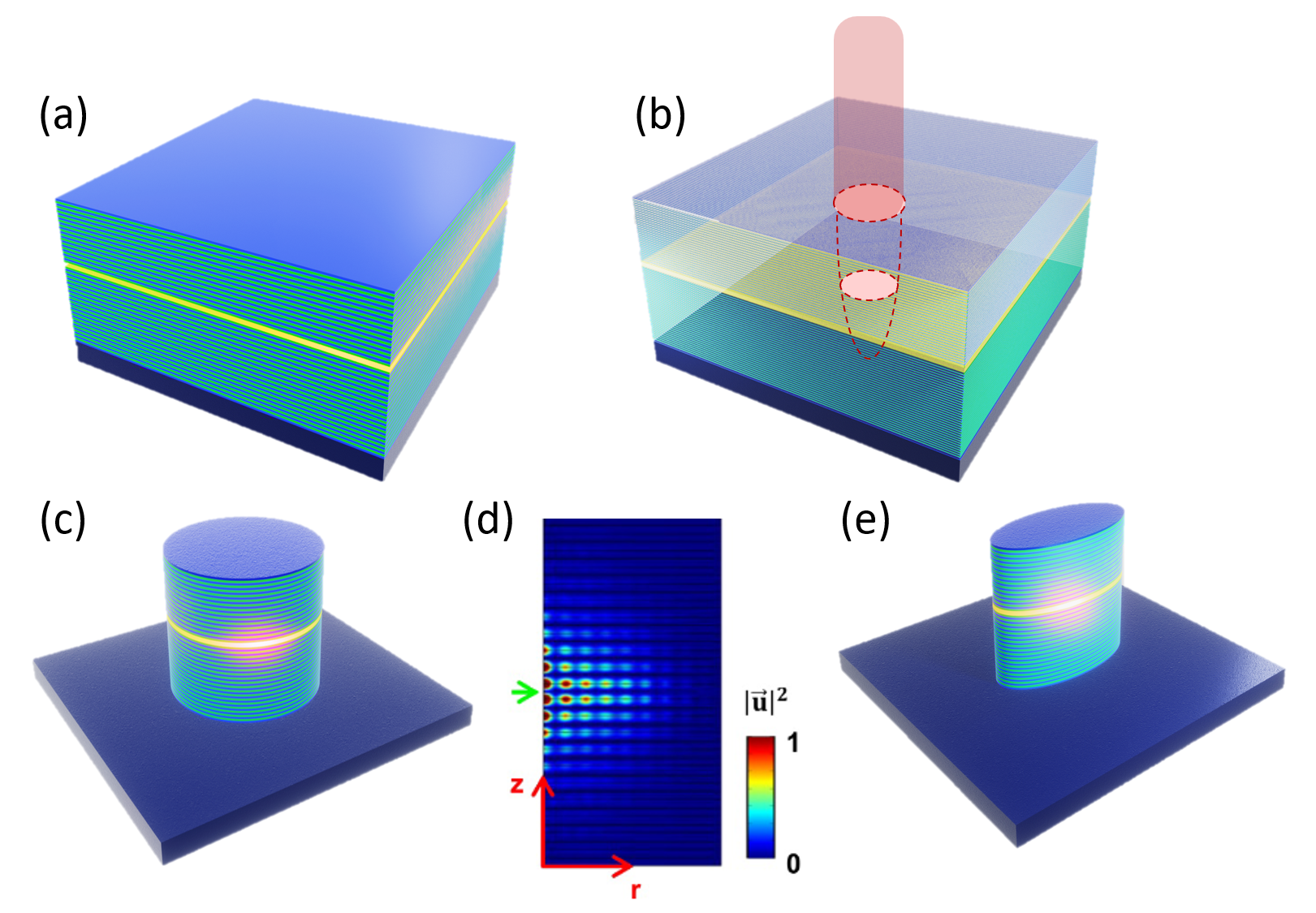}
    \caption{Architectures enabling three-dimensional confinement of acoustic and optical fields. (a) Planar optophononic microcavity. (b) Light-induced three-dimensional trapping scheme in planar structures. (c) Circular micropillar resonator providing lateral confinement through index and acoustic impedance contrast. (d) Calculated displacement profile of a confined acoustic mode in a circular micropillar.(adapted from \cite{lambertiMicropillar2017}) (e) Elliptical micropillar geometry enabling polarization-sensitive optophononic interactions. These architectures enable full three-dimensional confinement and form the basis for scalable and integrable nanophononic devices.}
    \label{architectures }
\end{figure*}

The development of GaAs/AlAs nanophononic devices critically relies on fabrication strategies that enable precise control of acoustic confinement, scalability across a single chip, and compatibility with hybrid optoelectronic platforms. In this context, bottom-up epitaxial growth techniques, combined with advanced nanofabrication, have played a central enabling role in transitioning from planar phononic structures to fully three-dimensional nanophononic architectures (Figure~\ref{architectures }).

Nanophononic mirrors and cavities based on GaAs/AlAs heterostructures are typically fabricated using molecular beam epitaxy (MBE) \cite{PanishMolecularBeamEpitaxy}. This technique provides atomic-scale control over layer thickness and interface quality, which is essential for operation in the ultrahigh-frequency regime, where acoustic wavelengths are on the order of a few nanometers. High structural quality minimizes phonon scattering and dephasing, enabling reproducible confinement at gigahertz and sub-terahertz frequencies. In addition, introducing a controlled thickness gradient during growth allows systematic tuning of cavity resonance frequencies across the wafer, facilitating spatial mapping and large-scale device characterization \cite{rodriguezFiberbasedAngularFiltering20to300GHz2021}.

Starting from planar multilayer structures that provide acoustic confinement along the growth direction, three-dimensional nanophononic architectures can be realized by patterning the heterostructures into micro- and nanostructures. In these geometries, vertical confinement is maintained by the DBRs, while lateral confinement arises from the strong contrast in optical and acoustic properties between the semiconductor material and the surrounding air. This additional degree of freedom enables enhanced control over both acoustic and optical mode profiles, as well as their mutual interaction.

Three-dimensional confinement is achieved by etching the epitaxially grown structures into micropillars or more complex geometries (Figure~\ref{architectures }c,e). High-resolution electron-beam lithography allows precise definition of lateral dimensions and shapes at the sub-100\,nm scale \cite{GrovesElectronBeamLithography,ReitzensteinMicropillarCavities}, while optical lithography provides scalable access to larger-area patterned structures using predefined masks \cite{dousse_controlled_2008}. These patterning steps are followed by dry etching techniques such as inductively coupled plasma (ICP) etching or reactive ion etching (RIE), which enable deep, anisotropic etching with high aspect ratios and well-controlled sidewall profiles \cite{Pearton_Norton_2004,Lee_2025_Etching}. Such capabilities are essential for achieving strong lateral confinement and reproducible device geometries.

These fabrication approaches allow the realization of circular micropillars, as well as more complex architectures such as embedded or elliptical pillars, where controlled ellipticity can be exploited for polarization-selective optophononic interactions \cite{rodriguezBrillouinEllipticalupillars2023}. In all cases, optical confinement arises from the DBRs in the vertical direction and from the refractive index contrast in the lateral direction, analogous to light guiding in optical fibers. The same fabrication strategies can be extended to define optophononic waveguides directly from planar cavity structures, enabling guided acoustic propagation and on-chip interference effects \cite{Xiang_Interference_Waveguide}.

Importantly, micropillar cavities fabricated using these methods are already well established in related fields, including exciton-polariton physics \cite{BajoniPolariton2008,KuriakosePolariton2022}, nonlinear optics \cite{Kuszelewicz_Optical_2000,Rivera_Reduced_1994}, and solid-state quantum optics, where quantum dots act as highly coherent artificial atoms embedded within micropillar resonators \cite{dousse_controlled_2008,Frey_ElectroOptic_2018}. The close technological overlap between these systems and GaAs/AlAs optophononic resonators provides a clear pathway for engineering hybrid devices in which confined acoustic modes interact with optical and electronic excitations, enabling advanced functionalities for integrated and quantum nanophononic platforms.

\section{High-frequency acoustic phonon dynamics}
\label{experiment}

\subsection{Planar heterostructures}

The confinement of acoustic phonons in planar GaAs/AlAs nanocavities strongly enhances optophononic interactions, in direct analogy with optical field enhancement in photonic cavities. Early experimental demonstrations established that embedding an acoustic cavity at the position of maximum optical confinement significantly increases phonon generation and detection efficiency \cite{trigoPlanarCavityPrl2002}. These studies built upon earlier investigations of phonon propagation in GaAs/AlAs superlattices, where Bragg reflection enabled measurements of phonon attenuation and velocity over a broad temperature range \cite{MarisGaAsAttenuationVelocitySuperLattice1994}. Together, these works demonstrated that planar heterostructures provide a robust and reproducible platform for accessing coherent high-frequency acoustic phonons.

\begin{figure*}
    \includegraphics[width=1\textwidth]{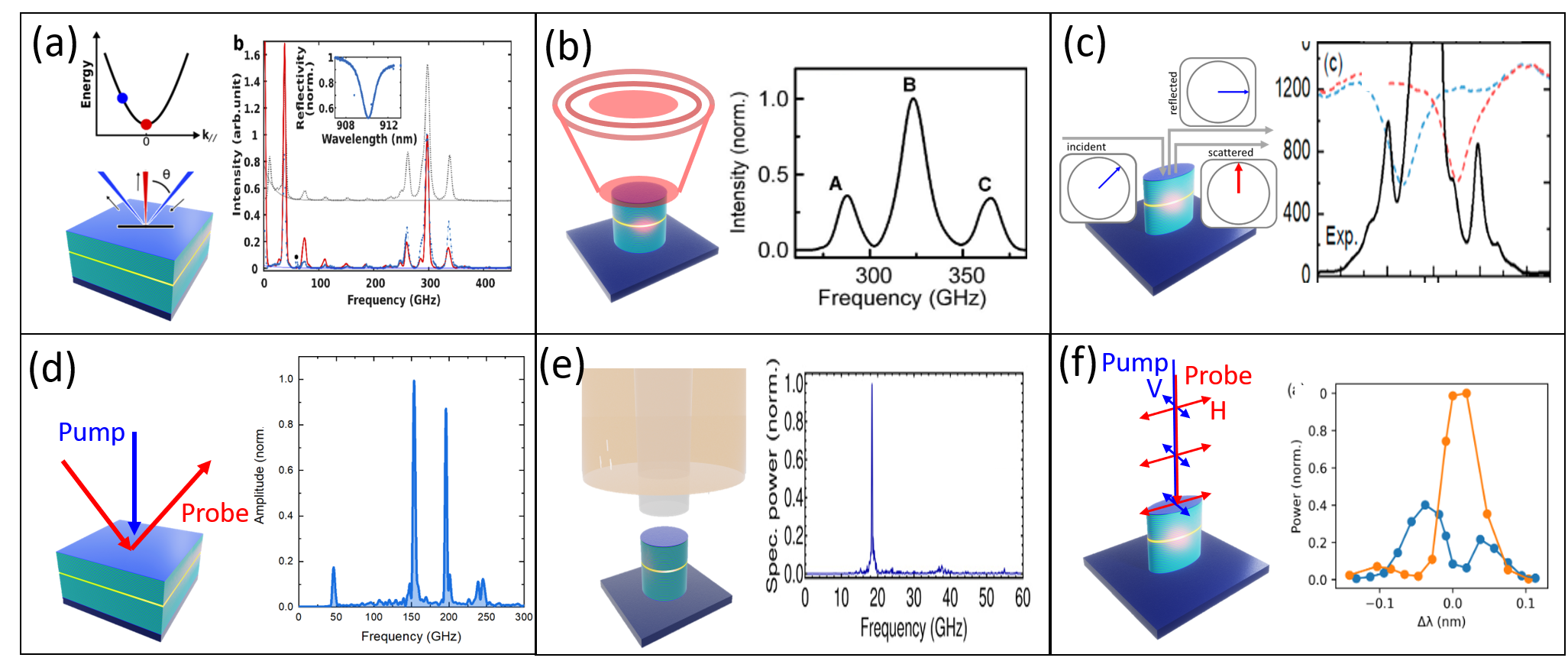}
    \caption{Experimental techniques for probing high-frequency acoustic phonons in the 20-300 GHz frequency range. Representative measurements demonstrating access to confined phonon modes in GaAs/AlAs optophononic resonators using Brillouin scattering and ultrafast pump–probe spectroscopy. (a–c) Brillouin scattering measurements: (a) double optical resonance (DOR) condition and Brillouin spectra of a planar cavity (adapted from \cite{rodriguezFiberbasedAngularFiltering20to300GHz2021}); (b) Spatial-filtering technique and Brillouin spectrum of a micropillar resonator revealing confined phonon modes \cite{esmannBrillouinScattering300GHz2019}; (c) polarization-dependent Brillouin scattering in an elliptical micropillar illustrating phonon selection rules \cite{rodriguezBrillouinEllipticalupillars2023}. (d–f) Ultrafast pump–probe measurements of coherent high-frequency phonons in planar, fiber-integrated micropillar \cite{ortizFiberintegratedMP}, and elliptical resonators \cite{Xiang2024_Micropillar_PumpProbe}.}
    \label{techniques}
\end{figure*}

Experimental access to confined phonon modes relies primarily on ultrafast optical techniques and complementary spectroscopic techniques (Figure~\ref{techniques}). Picosecond ultrasonics enables coherent phonon generation and detection through pump--probe modulation of optical properties, while asynchronous optical sampling (ASOPS) improves temporal stability and acquisition speed by eliminating mechanical delay lines \cite{Gebs2010_ASOPS,Velsink2023_ASOPS}. Complementary spectroscopic approaches based on Raman and Brillouin scattering provide non-invasive probes of phonon modes over a wide frequency range. Recent developments, including interferometric detection schemes and fiber-based angular filtering, have significantly improved signal-to-background ratios and enabled background-free detection of confined acoustic modes in planar nanocavities \cite{rodriguezFiberbasedAngularFiltering20to300GHz2021}.

A central challenge for functional nanophononics lies in achieving efficient and localized phonon generation without relying exclusively on optical confinement. Selective optical excitation of GaAs/AlAs heterostructures, achieved by tuning the pump wavelength to electronic resonances in quantum wells, has emerged as an effective strategy for enhancing phonon generation directly within acoustic cavities \cite{pascualSelectiveOpticalGenerationNanocavity,matsudaWavelengthSelectivePhotoexcitation2002,matsudaAcousticPhononGeneration2005}. This approach enables standalone excitation of confined phonon modes and provides increased flexibility for device integration.

In optophononic cavities, simultaneous confinement of optical and acoustic modes leads to strong photon--phonon coupling. GaAs/AlAs DBR-based microcavities are particularly well suited for this purpose, as comparable impedance and velocity contrasts for light and sound allow co-localization of electromagnetic and displacement fields with similar quality factors and spatial profiles. Further enhancement of optophononic coupling has been achieved using perturbed superlattices, band inversion strategies, and topological designs, enabling control over mode localization and robustness against disorder \cite{fainsteinSimultaneousConfinementPRL2013,LanzillottiKimura_TowardsGHzTHzcavity_2015,arreguiAndersonPhotonPhononColocalization2019,ortizTopologicalOpticalPhononic2021}.

Beyond single cavities, planar heterostructures allow the engineering of effective phonon potentials through coupled cavities or adiabatically modulated superlattices. These architectures support localized and extended phononic states, analogous to electronic minibands in solids, and enable controlled manipulation of phonon dispersion and group velocity \cite{kimuraPhononMolecules}. Quasiperiodic GaAs/AlAs superlattices have demonstrated long-lived confined phonon modes at room temperature, with frequencies approaching 100~GHz, and have revealed the potential for phonon-assisted control of carrier cooling dynamics \cite{Hanif_LongLived_2024}. Advanced excitation schemes based on subharmonic resonant optical driving further extend the accessible frequency range and spectral resolution, enabling precise measurements of phonon lifetimes and quality factors well beyond the limits of conventional pump--probe techniques \cite{bruchhausenSubharmonicResonantOpticalASOPS2011}.

\subsection{Micropillar structures}

Micropillar resonators provide full three-dimensional confinement of acoustic phonons and photons, resulting in modified phonon dispersion relations and enhanced local density of states. These properties make micropillars particularly attractive for coherent phonon control and active nanophononic functionalities. Electrically driven GaAs/AlAs superlattice micropillars have enabled sound amplification by stimulated emission of radiation (SASER), demonstrating hypersound gain and sustained phonon oscillations at frequencies up to several hundreds of gigahertz \cite{akimovkentSaserPrl2010,akimovKentSaserNatcom2013,akimovKentPhononChipSciRep2015}. Coupling between phonons and polariton Bose--Einstein condensates has further enabled self-oscillating mechanical states and coherent phonon generation mediated by hybrid light--matter excitations \cite{fainsteinPolaritonSaser2020,kuznetsovMicrocavityPhonoritonsCoherent2023a,carraro-haddadSolidstateContinuousTime2024,sesinGiantOptomechanicalCoupling2023}.

Theoretical modeling of acoustic confinement in superlattice-based micropillars has highlighted the role of surface boundary conditions in enhancing mode localization and quality factors, predicting acoustic Fano resonances with small mode volumes and strong optomechanical coupling \cite{garcia-sanchezTheoryAcousticConfinementSuperlattice2016}. Experimental studies have confirmed these predictions, revealing confined acoustic modes at frequencies ranging from tens to hundreds of gigahertz and demonstrating record-high quality-factor--frequency products approaching $10^{14}$ at room temperature \cite{anguianoMicropillarResonatorsOptomechanics2017,lambertiMicropillar2017,lagoinMicropillar2019}. These figures of merit indicate that micropillar nanophononic cavities are compatible with coherent control and hybrid quantum applications.

Micropillar geometries also enable advanced readout schemes for high-frequency phonons. Spatial, interferometric, and polarization-based filtering techniques have been developed to suppress elastic laser scattering and enable background-free Brillouin spectroscopy in the 20--300~GHz range \cite{esmannBrillouinScattering300GHz2019,rodriguezBrillouinEllipticalupillars2023,Mehdi_polarization_2024}. These approaches provide access to spontaneous phonon dynamics in nanoscale resonators and are compatible with tunable optophononic architectures.

Finally, strain-mediated coupling between neighboring pillar resonators has enabled the exploration of collective phononic effects, including mode hybridization and avoided crossings controlled by geometry and inter-pillar spacing \cite{weigStraincoupled2019}. Together with recent demonstrations of nanoscale quantum systems coupled to confined mechanical modes \cite{munschResonantDrivingSinglePhoton2017}, these results highlight the potential of micropillar-based nanophononic platforms for scalable phononic networks and hybrid quantum technologies.

\section{Integration of optical and acoustic fields} \label{integration}

\begin{figure*}
    \includegraphics[width=1\textwidth]{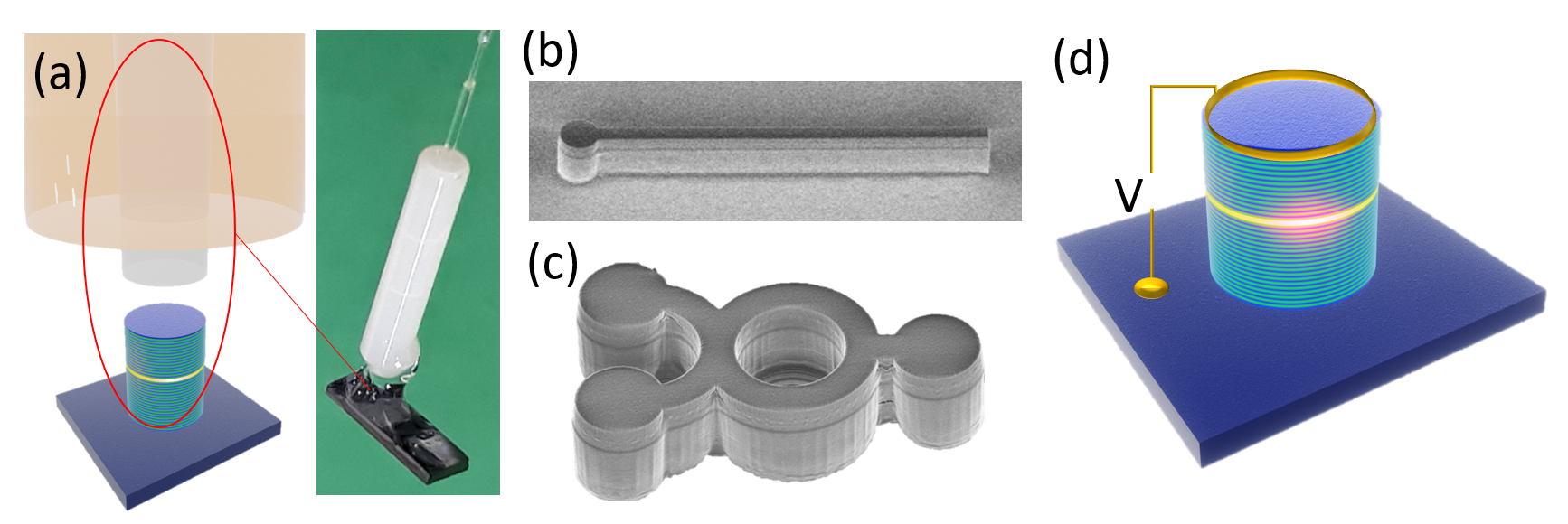}
    \caption{Integrated nanophononic platforms and routes toward functional devices.
(a) Micropillar optophononic resonators directly integrated with single-mode optical fibers for stable and efficient phonon generation and detection.~(adapted from \cite{ortizFiberintegratedMP}) (b,c) Scanning electron micrographs of micropillar-based waveguides and interferometric structures enabling guided phonon transport and phase control. (d) Schematic of optophononic resonators integrated with electrical contacts for on-chip and electrically driven phonon generation. These examples illustrate pathways toward scalable, hybrid, and multifunctional nanophononic circuits.}
    \label{integrated}
\end{figure*}

The small footprint of these nanophononic resonators, together with their ability to interact with optical and electronic excitations within the same material platform, makes them particularly attractive for integrated technologies (Figure~\ref{integrated}). The optical generation and detection of acoustic phonons in nanoacoustic micropillar resonators using pump-probe spectroscopy requires a precise mode matching between the pillar and the excitation/collection beam, which limits experiment stability and reproducibility. This challenge has been addressed by integrating single-mode fibers directly with the acoustic resonators (Figure~\ref{integrated}a). A permanent and stable mode overlap has been demonstrated by aligning the core of the fiber on top of the micropillar to ensure precise mode matching, and gluing it to the sample chip.~\cite{ortizFiberintegratedMP} Consequently, the elimination of the need for continuous realignment offers potential for consistent, plug-and-play experiments in individual microstructures.

Ultrahigh frequency acoustic phonons can also be exploited as information carriers, being a potential asset for communication systems and quantum technologies. The generation of a quasi-continuous source of coherent acoustic phonons at 20~GHz in nanocavity acoustic ridge waveguides has been recently achieved. In a non-local pump-probe configuration, where both excitation and detection beams are spatio-temporally separated, researchers detected acoustic signals at 20~$\mu$m away from the phonon source. The coherence of such waves was demonstrated by interfering two distant phonon sources and controlling the relative phase of the generated phonons.~\cite{Xiang_Interference_Waveguide} This achievement conveys the potential of acoustic phonons as information carriers, and constitutes one step forward on interfacing other solid-state platforms at ultrahigh frequencies.

To provide a compact benchmark view of the current platform status, Table~\ref{tab:performance_landscape} summarizes representative performance metrics, transduction approaches, and integration levels across the main GaAs/AlAs nanophononic architectures discussed above. In this context, these demonstrations indicate that nanophononic integration has progressed beyond isolated proof-of-concept devices. Planar and three-dimensional GaAs/AlAs architectures now support co-integrated phonon sources, resonators, waveguides, and readout schemes on a single chip, establishing a realistic pathway toward scalable nanophononic circuits compatible with optoelectronic and quantum technologies.

Beyond passive confinement, GaAs/AlAs nanophononic platforms now offer multiple and complementary control knobs that enable active functionality. Optical tuning, achieved through pump-power–dependent heating or carrier excitation, provides continuous and reversible control of phonon frequency and dissipation. Geometrical design parameters, including cavity thickness, lateral dimensions, and controlled ellipticity, allow deterministic tuning of mode spectra, polarization selectivity, and coupling strengths. Electrical control schemes, based on SASER operation, Stark-ladder superlattices, or integrated contacts, enable direct generation, amplification, and modulation of coherent acoustic phonons. In addition, coherent gigahertz acoustic waves can act as dynamic control fields for hybrid light–matter systems. Recent experiments have demonstrated that strain generated by bulk acoustic resonators can modulate the excitonic component of exciton–polariton states in semiconductor microcavities\cite{kuznetsovGroundstateExcitonPolariton2026}. This enables Floquet engineering of the polariton spectrum and facilitates selective population transfer between confined modes, thereby showcasing ultrafast acoustic control of optical and excitonic states. Together, these tuning mechanisms support key functional operations such as phase control, amplitude modulation, and switching, establishing GaAs/AlAs nanophononics as an actively reconfigurable materials platform rather than a static resonator system.

\begin{table}[!htbp]
\caption{Representative performance landscape of GaAs/AlAs nanophononic architectures.
The values are indicative and compiled from representative reports cited throughout this review.
Where applicable, ranges reflect typical operating conditions (temperature, geometry, and measurement technique).}
\label{tab:performance_landscape}

\footnotesize
\setlength{\tabcolsep}{2pt}
\renewcommand{\arraystretch}{1.7}

\begin{tabular}{llllll}

\parbox[t]{0.18\textwidth}{\raggedright\textbf{Architecture}} &
\parbox[t]{0.13\textwidth}{\raggedright\textbf{Typical frequency}} &
\parbox[t]{0.13\textwidth}{\raggedright\textbf{Representative performance metric}} &
\parbox[t]{0.13\textwidth}{\raggedright\textbf{Confinement}} &
\parbox[t]{0.18\textwidth}{\raggedright\textbf{Transduction / readout}} &
\parbox[t]{0.20\textwidth}{\raggedright\textbf{Integration level}} \\
\hline

\parbox[t]{0.18\textwidth}{\raggedright Planar DBR cavities \cite{trigoPlanarCavityPrl2002,MarisGaAsAttenuationVelocitySuperLattice1994,pascualSelectiveOpticalGenerationNanocavity,huynhSubterahertzPhononDynamics2006,Hanif_LongLived_2024,arreguiCoherentGenerationDetection2019,LanzillottiKimura_TowardsGHzTHzcavity_2015} }&
\parbox[t]{0.13\textwidth}{\raggedright $\sim$20--1000\,GHz (and higher orders)} &
\parbox[t]{0.13\textwidth}{\raggedright $Q \sim 10^{3}$--$10^{4}$;\; $Q\!\cdot\!f$ typically exceeding $10^{12}$}&
\parbox[t]{0.13\textwidth}{\raggedright 1D (vertical)} &
\parbox[t]{0.18\textwidth}{\raggedright Pump--probe;\; Raman/Brillouin;\; DOR-Brillouin} &
\parbox[t]{0.20\textwidth}{\raggedright On-wafer frequency mapping (growth gradient); compatible with photonic stacks} \\

\parbox[t]{0.18\textwidth}{\raggedright Optophononic planar microcavities \cite{fainsteinSimultaneousConfinementPRL2013, ortizTopologicalOpticalPhononic2021,arreguiAndersonPhotonPhononColocalization2019,rodriguezFiberbasedAngularFiltering20to300GHz2021,LanzillottiKimura_TowardsGHzTHzcavity_2015,kimuraRamanPrl2007}} &
\parbox[t]{0.13\textwidth}{\raggedright $\sim$20--600\,GHz} &
\parbox[t]{0.13\textwidth}{\raggedright Enhanced photon--phonon overlap; strong photoelastic coupling} &
\parbox[t]{0.13\textwidth}{\raggedright 1D (vertical) with optical co-localization} &
\parbox[t]{0.18\textwidth}{\raggedright Cavity-enhanced Brillouin;\; ultrafast spectroscopy} &
\parbox[t]{0.20\textwidth}{\raggedright Co-integrated optical and acoustic resonances; routes to polaritonic platforms} \\

\parbox[t]{0.18\textwidth}{\raggedright Micropillar optophononic resonators \cite{anguianoMicropillarResonatorsOptomechanics2017,lambertiMicropillar2017,lagoinMicropillar2019,esmannBrillouinScattering300GHz2019,rodriguezBrillouinEllipticalupillars2023,Mehdi_polarization_2024,Xiang2024_Micropillar_PumpProbe,ortizFiberintegratedMP}} &
\parbox[t]{0.13\textwidth}{\raggedright $\sim$20--300\,GHz (reported), extending upward} &
\parbox[t]{0.13\textwidth}{\raggedright Record $Q\!\cdot\!f$ up to $\sim 10^{14}$ at room temperature (representative)} &
\parbox[t]{0.13\textwidth}{\raggedright 3D (vertical + lateral)} &
\parbox[t]{0.18\textwidth}{\raggedright Brillouin (pattern/polarization filtering);\; pump--probe;\; interferometric readout} &
\parbox[t]{0.20\textwidth}{\raggedright Chip-scale arrays; fiber-coupled implementations; polarization-engineered designs} \\

\parbox[t]{0.18\textwidth}{\raggedright Guided nanophononic waveguides / ridges \cite{Xiang_Interference_Waveguide,yaremkevichProtectedLongDistanceGuiding2021,crespo-povedaGHzGuidedOptomechanics2022}} &
\parbox[t]{0.13\textwidth}{\raggedright $\sim$10--20\,GHz (demonstrated), scalable upward} &
\parbox[t]{0.13\textwidth}{\raggedright Propagation length / coherence set by scattering and coupling} &
\parbox[t]{0.13\textwidth}{\raggedright 1D--2D guided modes} &
\parbox[t]{0.18\textwidth}{\raggedright Non-local pump--probe; interferometric phase control} &
\parbox[t]{0.20\textwidth}{\raggedright On-chip routing, interference, and phase control of coherent phonons} \\

\parbox[t]{0.18\textwidth}{\raggedright Electrically driven active structures (SASER / Stark-ladder ; BAWRs) \cite{akimovkentSaserPrl2010,akimovKentSaserNatcom2013,akimovKentPhononChipSciRep2015,crespo-povedaGHzGuidedOptomechanics2022,kuznetsovMicrocavityPhonoritonsCoherent2023a,kuznetsovElectricallyDrivenMicrocavity2021}}&
\parbox[t]{0.13\textwidth}{\raggedright $\sim$10\,GHz--THz (device-dependent)} &
\parbox[t]{0.13\textwidth}{\raggedright Narrow linewidth / gain; electrically controlled intensity and phase} &
\parbox[t]{0.13\textwidth}{\raggedright Cavity / vertical structures} &
\parbox[t]{0.18\textwidth}{\raggedright Electrical injection; electro-acoustic transduction} &
\parbox[t]{0.20\textwidth}{\raggedright Toward compact sources and modulators; interface to electronics} \\

\parbox[t]{0.18\textwidth}{\raggedright Hybrid quantum nanomechanics (QD / polariton / spin-coupled)\cite{vlasovModernProblemsUltrafast2022,munschResonantDrivingSinglePhoton2017,TanosNanowire2024,krennerInterfacingQuantumEmitters2018,krennerOnchipGenerationDynamic2022,carraro-haddadSolidstateContinuousTime2024,kuznetsovGroundstateExcitonPolariton2026}} &
\parbox[t]{0.13\textwidth}{\raggedright sub-GHz -- GHz (demonstrated) with pathways upward} &
\parbox[t]{0.13\textwidth}{\raggedright Coherent coupling, occupancy control at cryogenic $T$} &
\parbox[t]{0.13\textwidth}{\raggedright Device-dependent (often 3D microstructures)} &
\parbox[t]{0.18\textwidth}{\raggedright Optical resonant driving; strain-mediated coupling; hybrid readout} &
\parbox[t]{0.20\textwidth}{\raggedright Pathways to hybrid quantum nodes and transducers} \\

\end{tabular}
\end{table}

\section{Conclusion and perspectives}
\label{conclusion}

GaAs/AlAs heterostructures have emerged as a mature and versatile platform for high-frequency nanophononics, enabling controlled confinement and manipulation of acoustic phonons from the gigahertz to the terahertz regime. The large acoustic impedance contrast, compatibility with state-of-the-art epitaxial growth, strong optoacoustic coupling, and quasi-linear phonon dispersion provide a robust materials framework for engineering nanophononic cavities with high quality factors, strong confinement, and efficient interaction with optical and electronic degrees of freedom. As highlighted throughout this Review, these properties have enabled steady progress from early proof-of-concept demonstrations toward functional and integrable nanophononic devices. Collectively, GaAs/AlAs nanophononic cavities now operate reproducibly from tens of gigahertz to the sub-terahertz regime, achieve quality-factor–frequency products exceeding $10^{12}$--$10^{14}$, and support chip-scale integration of optical, acoustic, and electrical functionalities.

Beyond linear phonon confinement, nonlinear acoustic phenomena offer promising opportunities to extend the functional landscape of nanophononics. In particular, acoustic solitons in GaAs-based systems provide a unique route to generate ultrashort, high-frequency strain pulses with nanometer-scale spatial extent and picosecond temporal resolution. These nonlinear excitations enable the probing of vibrational, thermal, and transport properties of nanostructures under extreme conditions, and open perspectives for exploring ultrafast and non-equilibrium phonon dynamics at the nanoscale. The controlled generation and detection of KdV-type acoustic solitons therefore represent an important avenue for future developments in ultrafast acoustics and phonon-based material characterization \cite{peronneAcousticSolitonsRobust2017,peronneAcousticSolitons2006}.

Another emerging direction concerns the extension of nanophononic networks and coupled architectures toward higher frequencies. While arrays of acoustic resonators operating in the MHz regime have enabled the exploration of collective dynamics, energy transport, and topological effects, their translation to the GHz--THz domain remains largely unexplored. Accessing these higher frequencies is not only attractive for faster signal modulation, but also for reaching low phonon occupation numbers at standard cryogenic temperatures. This regime is particularly relevant for quantum technologies, where coherent nanophononic networks could be specifically tailored for quantum communication and information processing applications.

Hybrid nanophononic systems constitute a further key perspective enabled by GaAs/AlAs acoustic nanocavities. The ability to co-confine high-frequency phonons with optical modes and quantum emitters opens the door to hybrid optomechanical platforms in which acoustic excitations mediate interactions between photons, excitons, spins, or other solid-state quantum systems. Such hybrid architectures are especially promising for interfacing localized quantum emitters with confined or propagating acoustic modes, taking advantage of the strong and versatile coupling of strain to electronic and optical excitations in semiconductors \cite{TanosNanowire2024,KrennerHybridGaAsLiNbO3Surface2020,krennerInterfacingQuantumEmitters2018,krennerOnchipGenerationDynamic2022}.

Despite these advances, efficient and scalable transduction of high-frequency acoustic phonons remains a central challenge for practical applications. Optical generation and detection schemes, while powerful, typically rely on bulky and high-power laser systems. Recent progress in electrically driven phonon generation, using approaches such as Stark-ladder superlattices or bulk acoustic wave generators, provides promising routes toward compact and energy-efficient nanophononic devices \cite{kuznetsovElectricallyDrivenMicrocavity2021,akimovkentSaserPrl2010}. In this context, exploiting quantum transport effects in resonant tunneling structures for electro-acoustic transduction could help bridge the gap between electronic devices and coherent acoustic phonons \cite{cardozodeoliveiraElectroluminescenceOnoffRatio2018,cardozodeoliveiraDeterminationCarrierDensity2021}.

Finally, the interaction of ultrafast acoustic phonons with magnetic degrees of freedom offers additional opportunities for functional nanophononics. Recent developments in ultrafast magnetoacoustics demonstrate that picosecond strain pulses can efficiently manipulate magnetization dynamics in magnetic films and semiconductors, enabling control over ferromagnetic resonance, magnetization switching, and nonlinear spin dynamics \cite{vlasovModernProblemsUltrafast2022}. Furthermore, picosecond strain pulses have proven effective for probing magnetization in ferromagnetic semiconductors such as GaMnAs, opening new perspectives for ultrafast magnetic switching and strain-mediated spin control \cite{thevenardEffectPicosecondStrain2010}.

From a near-term perspective, progress over the next five to ten years is expected to focus on deterministic integration of quantum emitters, improved epitaxial growth combining GaAs/AlAs with complementary material systems, and continued advances in high-resolution lithography enabling smaller device footprints and increased architectural complexity. In parallel, efficient electrically driven phonon generation and detection will be essential to reduce reliance on ultrafast optical systems and to enable compact, low-power, and scalable nanophononic components. These developments are expected to support reproducible nanophononic cavities operating well into the sub-terahertz and terahertz regimes, while facilitating the implementation of on-chip circuits combining phonon sources, waveguides, and resonators, as well as small-scale coupled networks compatible with hybrid quantum architectures.

 In the quantum regime, operation at standard cryogenic temperatures is expected to enable access to low phonon occupation numbers, making GaAs/AlAs nanophononic platforms compatible with hybrid quantum architectures involving quantum dots, exciton-polaritons, or spin-based systems. Together, these developments would establish nanophononics as a functional layer within integrated optoelectronic and quantum technologies, rather than a collection of standalone laboratory demonstrations. Finally, operation in the quantum regime constitutes a longer-term but well-defined challenge, requiring simultaneous advances in dissipation control, phonon occupation, and coherent coupling to quantum emitters. Addressing these challenges in a staged manner—from active control, to electrical interfacing, and ultimately to quantum operation—provides a realistic roadmap for transforming nanophononic cavities into functional building blocks of future hybrid technologies.
 
 The next milestone for GaAs/AlAs nanophononics is the transition from high-quality resonators to fully integrated phononic systems, where confined acoustic modes provide active functionality for information processing, transduction, and hybrid quantum control.

\medskip
\textbf{Acknowledgements} \par 
The authors acknowledge financial support from the European Research Council (ERC) under the European Union’s Horizon Europe research and innovation programme (Grant No. 101045089, T-Recs).
The authors also thank A. Lemaître, I. Sagnes, A. Harouri, and M. Esmann for their valuable discussions and for providing the device images used in Figure 5.

\bibliographystyle{unsrt}
\bibliography{references}

\end{document}